\documentclass{ws-procs9x6-cpt22}
\begin{document}

\def\al{\alpha}
\def\be{\beta}
\def\ga{\gamma}
\def\de{\delta}
\def\ep{\epsilon}
\def\ve{\varepsilon}
\def\ze{\zeta}
\def\et{\eta}
\def\th{\theta}
\def\vt{\vartheta}
\def\io{\iota}
\def\vka{\varkappa}
\def\ka{\kappa}
\def\la{\lambda}
\def\vpi{\varpi}
\def\rh{\rho}
\def\vr{\varrho}
\def\si{\sigma}
\def\vs{\varsigma}
\def\ta{\tau}
\def\up{\upsilon}
\def\ph{\phi}
\def\vp{\varphi}
\def\ch{\chi}
\def\ps{\psi}
\def\om{\omega}
\def\Ga{\Gamma}
\def\De{\Delta}
\def\Th{\Theta}
\def\La{\Lambda}
\def\Si{\Sigma}
\def\Up{\Upsilon}
\def\Ph{\Phi}
\def\Ps{\Psi}
\def\Om{\Omega}
\def\cA{{\cal A}}
\def\cB{{\cal B}}
\def\cC{{\cal C}}
\def\cD{{\cal D}}
\def\cE{{\cal E}}
\def\cH{{\cal H}}
\def\cl{{\cal L}}
\def\cL{{\cal L}}
\def\cO{{\cal O}}
\def\cV{{\cal V}}
\def\cP{{\cal P}}
\def\cR{{\cal R}}
\def\cS{{\cal S}}
\def\cT{{\cal T}}

\def\fr#1#2{{{#1}\over{#2}}}
\def\frac#1#2{{\textstyle{{#1}\over{#2}}}}
\def\half{{\textstyle{1\over 2}}}
\def\ol{\overline}
\def\prt{\partial}
\def\pt{\phantom}

\def\Re{\hbox{Re}\,}
\def\Im{\hbox{Im}\,}

\def\lsim{\mathrel{\rlap{\lower4pt\hbox{\hskip1pt$\sim$}}
    \raise1pt\hbox{$<$}}}
\def\gsim{\mathrel{\rlap{\lower4pt\hbox{\hskip1pt$\sim$}}
    \raise1pt\hbox{$>$}}}

\def\etal{{\it et al.}}
\def\nn{\nonumber}

\def\vev#1{\langle {#1}\rangle}
\def\expect#1{\langle{#1}\rangle}
\def\bra#1{\langle{#1}|}
\def\ket#1{|{#1}\rangle}

\def\tr{{\rm tr}}

\newcommand{\beq}{\begin{equation}}
\newcommand{\eeq}{\end{equation}}
\newcommand{\bea}{\begin{eqnarray}}
\newcommand{\eea}{\end{eqnarray}}
\newcommand{\rf}[1]{(\ref{#1})}
\newcommand{\eq}[1]{Eq.~(\ref{#1})}
\newcommand{\Eq}[1]{Equation (\ref{#1})}
\newcommand{\chp}[1]{Chap.~\ref{#1}}
\newcommand{\Chp}[1]{Chapter~\ref{#1}}
\newcommand{\Tab}[1]{Table~\ref{#1}}
\newcommand{\Sec}[1]{Section~\ref{#1}}
\newcommand{\Subsec}[1]{Subsection~\ref{#1}}

\def\psb{\ol\ps{}}
\def\mbf#1{\boldsymbol #1}

\def\pvec{\mbf p}
\def\gavec{\mbf\ga}

\def\Q{\mathcal Q}
\def\S{\mathcal S}
\def\P{\mathcal P}
\def\V{\mathcal V}
\def\A{\mathcal A}
\def\T{\mathcal T}
\def\Qhat{\widehat\Q}
\def\Shat{\widehat\S}
\def\Phat{\widehat\P}
\def\Vhat{\widehat\V}
\def\Ahat{\widehat\A}
\def\That{\widehat\T}

\def\X{X}
\def\Y{Y}
\def\Z{Z}
\def\Xhat{\widehat\X}
\def\Yhat{\widehat\Y}
\def\Zhat{\widehat\Z}

\def\codt{\cos{\om_\oplus T_\oplus}}
\def\sodt{\sin{\om_\oplus T_\oplus}}
\def\ctodt{\cos{2\om_\oplus T_\oplus}}
\def\stodt{\sin{2\om_\oplus T_\oplus}}
\def\cthodt{\cos{3\om_\oplus T_\oplus}}
\def\sthodt{\sin{3\om_\oplus T_\oplus}}

\def\cmtemplate#1#2#3#4{{#1}^{#3}_{#4}}
\def\mfcm#1#2{\cmtemplate{m}{#1}{#2}{5}}
\def\acm#1#2{\cmtemplate{a}{#1}{#2}{}}
\def\bcm#1#2{\cmtemplate{b}{#1}{#2}{}}
\def\ccm#1#2{\cmtemplate{c}{#1}{#2}{}}
\def\dcm#1#2{\cmtemplate{d}{#1}{#2}{}}
\def\ecm#1#2{\cmtemplate{e}{#1}{#2}{}}
\def\fcm#1#2{\cmtemplate{f}{#1}{#2}{}}
\def\gcm#1#2{\cmtemplate{g}{#1}{#2}{}}
\def\Hcm#1#2{\cmtemplate{H}{#1}{#2}{}}

\def\ctemplate#1#2#3#4{{#1}^{(#2)#3}_{#4}}
\def\mc#1#2{\ctemplate{m}{#1}{#2}{}}
\def\mfc#1#2{\ctemplate{m}{#1}{#2}{5}}
\def\ac#1#2{\ctemplate{a}{#1}{#2}{}}
\def\bc#1#2{\ctemplate{b}{#1}{#2}{}}
\def\cc#1#2{\ctemplate{c}{#1}{#2}{}}
\def\dc#1#2{\ctemplate{d}{#1}{#2}{}}
\def\ec#1#2{\ctemplate{e}{#1}{#2}{}}
\def\fc#1#2{\ctemplate{f}{#1}{#2}{}}
\def\gc#1#2{\ctemplate{g}{#1}{#2}{}}
\def\Hc#1#2{\ctemplate{H}{#1}{#2}{}}

\def\mcf#1#2{\ctemplate{m}{#1}{#2}{F}}
\def\mfcf#1#2{\ctemplate{m}{#1}{#2}{5F}}
\def\acf#1#2{\ctemplate{a}{#1}{#2}{F}}
\def\bcf#1#2{\ctemplate{b}{#1}{#2}{F}}
\def\ccf#1#2{\ctemplate{c}{#1}{#2}{F}}
\def\dcf#1#2{\ctemplate{d}{#1}{#2}{F}}
\def\ecf#1#2{\ctemplate{e}{#1}{#2}{F}}
\def\fcf#1#2{\ctemplate{f}{#1}{#2}{F}}
\def\gcf#1#2{\ctemplate{g}{#1}{#2}{F}}
\def\Hcf#1#2{\ctemplate{H}{#1}{#2}{F}}

\def\mcpf#1#2{\ctemplate{m}{#1}{#2}{\prt F}}
\def\mfcpf#1#2{\ctemplate{m}{#1}{#2}{5\prt F}}
\def\acpf#1#2{\ctemplate{a}{#1}{#2}{\prt F}}
\def\bcpf#1#2{\ctemplate{b}{#1}{#2}{\prt F}}
\def\ccpf#1#2{\ctemplate{c}{#1}{#2}{\prt F}}
\def\dcpf#1#2{\ctemplate{d}{#1}{#2}{\prt F}}
\def\ecpf#1#2{\ctemplate{e}{#1}{#2}{\prt F}}
\def\fcpf#1#2{\ctemplate{f}{#1}{#2}{\prt F}}
\def\gcpf#1#2{\ctemplate{g}{#1}{#2}{\prt F}}
\def\Hcpf#1#2{\ctemplate{H}{#1}{#2}{\prt F}}

\def\cmtemplate#1#2#3#4{{#1}^{#3}_{#4}}
\def\mfcmw#1#2#3{\cmtemplate{m}{#1}{#2}{5{,#3}}}
\def\acmw#1#2#3{\cmtemplate{a}{#1}{#2}{{#3}}}
\def\bcmw#1#2#3{\cmtemplate{b}{#1}{#2}{{#3}}}
\def\ccmw#1#2#3{\cmtemplate{c}{#1}{#2}{{#3}}}
\def\dcmw#1#2#3{\cmtemplate{d}{#1}{#2}{{#3}}}
\def\ecmw#1#2#3{\cmtemplate{e}{#1}{#2}{{#3}}}
\def\fcmw#1#2#3{\cmtemplate{f}{#1}{#2}{{#3}}}
\def\gcmw#1#2#3{\cmtemplate{g}{#1}{#2}{{#3}}}
\def\Hcmw#1#2#3{\cmtemplate{H}{#1}{#2}{{#3}}}

\def\ctemplate#1#2#3#4{{#1}^{(#2)#3}_{#4}}
\def\mcw#1#2#3{\ctemplate{m}{#1}{#2}{{#3}}}
\def\mfcw#1#2#3{\ctemplate{m}{#1}{#2}{5{,#3}}}
\def\acw#1#2#3{\ctemplate{a}{#1}{#2}{{#3}}}
\def\bcw#1#2#3{\ctemplate{b}{#1}{#2}{{#3}}}
\def\ccw#1#2#3{\ctemplate{c}{#1}{#2}{{#3}}}
\def\dcw#1#2#3{\ctemplate{d}{#1}{#2}{{#3}}}
\def\ecw#1#2#3{\ctemplate{e}{#1}{#2}{{#3}}}
\def\fcw#1#2#3{\ctemplate{f}{#1}{#2}{{#3}}}
\def\gcw#1#2#3{\ctemplate{g}{#1}{#2}{{#3}}}
\def\Hcw#1#2#3{\ctemplate{H}{#1}{#2}{{#3}}}

\def\mcfw#1#2#3{\ctemplate{m}{#1}{#2}{F{,#3}}}
\def\mfcfw#1#2#3{\ctemplate{m}{#1}{#2}{5F{,#3}}}
\def\acfw#1#2#3{\ctemplate{a}{#1}{#2}{F{,#3}}}
\def\bcfw#1#2#3{\ctemplate{b}{#1}{#2}{F{,#3}}}
\def\ccfw#1#2#3{\ctemplate{c}{#1}{#2}{F{,#3}}}
\def\dcfw#1#2#3{\ctemplate{d}{#1}{#2}{F{,#3}}}
\def\ecfw#1#2#3{\ctemplate{e}{#1}{#2}{F{,#3}}}
\def\fcfw#1#2#3{\ctemplate{f}{#1}{#2}{F{,#3}}}
\def\gcfw#1#2#3{\ctemplate{g}{#1}{#2}{F{,#3}}}
\def\Hcfw#1#2#3{\ctemplate{H}{#1}{#2}{F{,#3}}}

\def\mn{{\mu\nu}}
\def\ma{{\mu\al}}
\def\mna{{\mu\nu\al}}
\def\ab{{\al\be}}
\def\bec{{\be\ga}}
\def\mab{{\mu\al\be}}
\def\mnab{{\mu\nu\al\be}}
\def\abc{{\al\be\ga}}
\def\bca{{\be\ga\al}}
\def\cab{{\ga\al\be}}
\def\mabc{{\mu\al\be\ga}}
\def\mnabc{{\mu\nu\al\be\ga}}
\def\abcd{{\al\be\ga\de}}
\def\va{{\vs\al}}
\def\vab{{\vs\al\be}}
\def\vabc{{\vs\al\be\ga}}
\def\vabcd{{\vs\al\be\ga\de}}

\def\m{m_\ps}
\def\mw{m_w}

\def\quar{\frac 1 4}

\def\En{E_{n, s}}
\def\Epn{E_{n,\pm1}^{e^+}}
\def\app{\approx}
\def\cthodt{\cos{3\om_\oplus T_\oplus}}
\def\sthodt{\sin{3\om_\oplus T_\oplus}}
\def\note#1{{\it note \cite{#1}}}
\def\ens{E_{n,\pm}}
\def\enms{E_{n,-s}}
\def\at{\widetilde a}
\def\bt{\widetilde b}
\def\bft{{\widetilde b}_F}
\def\mft{{\widetilde m}_F}
\def\atw#1#2{{\widetilde a}_{#1}^{#2}}
\def\btw#1#2{{\widetilde b}_{#1}^{#2}}
\def\bftw#1#2{{\widetilde b}_{F,#1}^{#2}}
\def\mftw#1#2{{\widetilde m}_{F,#1}^{#2}}
\def\atws#1#2{{\widetilde a}_{#1}^{*#2}}
\def\btws#1#2{{\widetilde b}_{#1}^{*#2}}
\def\bftws#1#2{{\widetilde b}_{F,#1}^{*#2}}
\def\mftws#1#2{{\widetilde m}_{F,#1}^{*#2}}

\def\bptw#1#2{{\widetilde b}_{#1}^{\prime #2}}
\def\mtw#1#2{{\widetilde m}_{#1}^{#2}}
\def\ctw#1#2{{\widetilde c}_{#1}^{#2}}
\def\bptws#1#2{{\widetilde b}_{#1}^{\prime *#2}}
\def\mtws#1#2{{\widetilde m}_{#1}^{*#2}}
\def\ctws#1#2{{\widetilde c}_{#1}^{*#2}}

\def\vos{\mathrel{\rlap{\lower0pt\hbox{\hskip0.5pt{$\scriptstyle s$}}}
    \raise2pt\hbox{$\scriptstyle \neg$}}}

\title{Searches for Lorentz and CPT Violation with Confined Particles}

\author{Yunhua Ding $^{1,2}$}

\address{$^1$Department of Physics and Astronomy, 
Ohio Wesleyan University, \\
Delaware, OH 43015, USA}

\address{$^2$W.M.\ Keck Science Department,
Claremont McKenna, Pitzer, and Scripps Colleges, \\
Claremont, CA 91711, USA}

\begin{abstract}
An overview of recent progress on searches for Lorentz- and CPT-violating signals 
with confined particles and antiparticles in Penning traps is presented. 
In the context of the Standard-Model Extension (SME),
leading-order shifts in the cyclotron and anomaly frequencies of a confined particle 
and antiparticle due to Lorentz and CPT violation are provided.
The two frequencies are then related to comparisons of 
charge-to-mass ratios and magnetic moments between particles and antiparticles.
Applying reported results from Penning-trap experiments 
leads to new limits on various coefficients for Lorentz violation. 
\end{abstract}

\bodymatter

\section{Introduction}
Penning traps have in recent years provided a powerful tool to 
precisely measure and compare fundamental properties between particles and antiparticles. 
For example,
the comparison of charge-to-mass ratios between protons and antiprotons has
recently reached a precision of 16 ppt by the BASE collaboration at CERN. \cite{22bo}
The same group has also measured the magnetic moments of protons and antiprotons 
to a precision of 0.3 ppb and 1.5 ppb, respectively. \cite{17sc, 17sm}
The impressive high sensitivities achieved by Penning-trap experiments 
offer an excellent opportunity to test fundamental symmetries 
of Einstein's General Relativity and the Standard Model of particle physics,
such as Lorentz and CPT symmetry. 
In recent years,
it has been shown that tiny violations of 
Lorentz and CPT symmetry can appear 
in a more fundamental theory of quantum gravity, 
such as strings.\cite{string}
The comprehensive framework to study all possible Lorentz violation is
the Standard-Model Extension (SME), \cite{SME}
constructed by adding all possible Lorentz-violating terms to the action of
General Relativity and the Standard Model. 
Each Lorentz-violating term in the SME is formed by contracting a Lorentz-violating operator
with a corresponding coefficient that controls the size of Lorentz violation.
The subset of the SME that restricts to operators of mass dimensions $d\leq 4$ 
is called the minimal SME,
while the nonminimal SME focuses on operators of mass dimensions $d>4$,
which can be viewed as higher-order corrections to the minimal SME.
In a general effective field theory, 
any CPT violation comes with violation in Lorentz symmetry, \cite{ck,owg} 
so the SME also characterizes all possible CPT violation. 
Over the past few decades, 
high-precision experiments across a large range of subfields of physics,
including Penning traps, 
have been performed to provide striking limits 
on coefficients for Lorentz violation in different SME sectors. \cite{datatables}

\section{Theory}
In a Lorentz-invariant case, 
for a single charged Dirac fermion of species $w$, charge $q$, and mass $m_w$ 
confined in a Penning trap with a magnetic field strength $B$,
its charge-to-mass ratio and $g$ factor (which is directly related to the magnetic moment)
are related to two frequencies, 
the cyclotron frequency $\om_c$ and the Larmor spin precession frequency $\om_L$,
given by
\bea
\dfrac{|q|}{m_w}   = \dfrac{\om_c}{B} ,
\quad
\fr {g}{2} = \fr {\om_L}{\om_c}
=
1+ \fr {\om_a}{\om_c},
\eea
where the difference $\om_a=\om_L-\om_c$ is the anomaly frequency.
In the presence of Lorentz and CPT violation,
the energy levels of the fermion could be modified due to contributions from Lorentz-violating operators.
As a result,
both the cyclotron frequency $\om_c$ and anomaly frequency $\om_a$ could be shifted.
In a reference frame with its positive $z$ axis chosen to be aligned with
the direction of the magnetic field used in the trap,
the leading-order contributions to $\om_a$ and $\om_c$ from Lorentz and CPT violation 
are given by 
\bea
\de \omega_c^{w}
&=&
\big(\dfrac{1}{m_w^2} \bptw w {3} 
- \dfrac{1}{m_w} (\ctw w {00} + \ctw w {11} + \ctw w {22})
- (\btw w {311} + \btw w {322})\big) |q|B ,
\nn\\
\de \om_a^{w} 
&=&
 2 \btw w 3 - 2 \bftw w {33} B ,
\eea
where the tilde coefficients are
different combinations of the fundamental coefficients for Lorentz violation,
defined in Refs.\ \refcite{20dr} and \refcite{16dk}.
The corresponding results of frequency shifts for an antifermion $\ol{w}$ 
can be obtained by reversing the signs of all the CPT-odd coefficients 
for Lorentz violation in the definitions of the tilde quantities.

We note that obtaining the frequency shift results in the Sun-centered frame,
the standard canonical frame where
the coefficients for Lorentz violation are assumed to be constant,
requires a transformation matrix involving in general 
the sidereal frequency of the Earth, the local sidereal time, 
the colatitude of the laboratory,
and a suitable set of Euler angles. \cite{suncenter, 20dr}

\section{Applications}
In a Penning-trap experiment,
the comparison of the charge-to-mass ratios between a particle and an antiparticle
is related to the difference in the shifts of their cyclotron frequencies by
\bea
\label{ratio-lv}
\dfrac{(|q|/m)_{\ol{w}}}{(|q|/m)_{w}} - 1
\longleftrightarrow
\dfrac{\om_c^{\ol{w}}}{\om_c^{w}} - 1
=
\dfrac{\de \om_c^{\ol{w}} -  \de \om_c^{w}} {\om_c^{w}}.
\eea
From the above result,
the tilde coefficients for Lorentz violation that are relevant to the comparison are
$\bptw w {3}$, 
$\ctw w {11} + \ctw w {22}$,
$\btw w {311} + \btw w {322}$,
$\bptws w {3}$, 
$\ctws w {11} + \ctws w {22}$,
and
$\btws w {311} + \btws w {322}$.
For the proton and antiproton charge-to-mass ratio comparison, 
the BASE collaboration achieved a precision of 69 ppt in 2015. \cite{15ul}
Applying the transformation matrix relevant to this experiment, 
the reported result was used to extract limits on various coefficients for Lorentz violation,
which were summarized in Table~1 and Table~2 in Ref.\ \refcite{21d}.
We note that very recently the BASE collaboration improved the comparison 
to a record precision of 16 ppt. \cite{22bo}
Therefore, 
an improvement of a factor of 4 for the constraints on the coefficients listed in 
Table~1 and Table~2 in Ref.\ \refcite{21d} is expected. 
While for the electron-positron charge-to-mass ratio comparison,  
a group at the University of Washington
reached a precision of 130 ppb \cite{81sc}. 
Following a similar analysis as in the proton sector, 
the reported precision leads to various limits on the coefficients for Lorentz violation
in the electron sector, 
listed in Table 3 in Ref.\ \refcite{21d}.

The particle-antiparticle $g$ factor comparisons 
are related to the shifts in both the cyclotron and anomaly frequencies, 
given by
\bea
\label{gratio-lv}
\half (g_w - g_{\ol w})
\longleftrightarrow
\fr {\om_a^w}{\om_c^w} - \fr {\om_a^{\ol w}}{\om_c^{\ol w}} =
\fr {\de\om_a^w}{\om_c^w} - \fr {\de\om_a^{\ol w}} {\om_c^{\ol w}}.
\eea 
This shows that the corresponding tilde coefficients for Lorentz violation are
$\btw w 3$, $\btws w 3$, $\bftw w {33}$, and~$\bftws w {33}$. 
In the proton sector,
the current best measurement of the magnetic moment for the proton and antiproton
were both obtained by the BASE collaboration,
with a sensitivity of 0.3 ppb \cite{17sc} and 1.5 ppb \cite{17sm}, respectively. 
Combining the two reported results and applying the relevant transformation matrices,
another set of tilde coefficients for Lorentz violation was constrained,
included in Table 4 in Ref. \refcite{21d}.
In the electron sector, 
the comparison of the anomaly frequencies between electrons and positrons 
was carried out at the University of Washington, 
with a precision of 2 ppt. \cite{de99}
The results of the limits on the tilde coefficients for Lorentz violation from this experiment 
were presented in Table 5 in Ref.~\refcite{21d}. 
Also, 
a sidereal-variation analysis of the electron anomaly frequencies was also 
carried out at Harvard University. \cite{07ha}
This gives constraints on additional coefficients for Lorentz violation
that cannot be accessed by a direct anomaly frequency comparison. 
The related results were also listed in Table 5 in Ref. \refcite{21d}. 

In summary,
Penning-trap experiments provide highly precise measurements 
of fundamental properties of confined particles to test Lorentz and CPT symmetry
and offer excellent coverage of the coefficients for Lorentz violation.
They continue to provide strong motivations to search for possible 
Lorentz- and CPT-violating signals.

\section*{Acknowledgments}
This work was supported in part by the W.M.\ Keck Science Department at
Claremont McKenna, Pitzer, and Scripps Colleges.

\end{document}